\documentclass[aip,jap,twocolumn,reprint]{revtex4-1}
\usepackage{amsmath}
\usepackage{amssymb}
\usepackage{graphicx}
\usepackage{hyperref}
\begin{document}
\title{Magnetism and electronic structure of ($001$)- and ($111$)-oriented LaTiO$_3$ bilayers sandwiched in LaScO$_3$ barriers}
\author{Yakui Weng}
\author{Shuai Dong}
\email{sdong@seu.edu.cn}
\affiliation{Department of Physics, Southeast University, Nanjing 211189, China}
\date{\today}

\begin{abstract}
In this study, the magnetism and electronic structure of LaTiO$_3$ bilayers along both the ($001$) and ($111$) orientations are calculated using the density functional theory. The band insulator LaScO$_3$ is chosen as the barrier layer and substrate to obtain the isolating LaTiO$_3$ bilayer. For both the ($001$)- and ($111$)-oriented cases, LaTiO$_3$ demonstrates the G-type antiferromagnetism as the ground state, similar to the bulk material. However, the electronic structure is significantly changed. The occupied bands of Ti are much narrower in the ($111$) case, giving a nearly flat band. As a result, the exchange coupling between nearest-neighbor Ti ions are reformed in these superlattices, which will affect the N\'{e}el temperature significantly.
\end{abstract}
\maketitle

In recent years, oxide heterostructures attracted many research attentions due to their plenty physics and promising potential for new electronic devices.\cite{Dagotto:Sci07} For example, the two-dimensional electronic gas ($2$DEG) was observed at the interfaces between insulating perovskites (e.g. SrTiO$_3$/LaTiO$_3$ and SrTiO$_3$/LaAlO$_3$),\cite{Ohtomo:Nat,Ohtomo:Nat04} and the metal-insulator transition was found in LaMnO$_3$/SrMnO$_3$ superlattices (SLs).\cite{Bhattacharya:Prl,Dong:Prb08.3} In addition, the electric-controllable magnetism was achieved in multiferroic heterostructures.\cite{Wu:Nm10,Dong:Prl2,Dong:Prb11.2,Huang:Mplb}

In particular, recent theoretical studies on multilayered structures predicted intriguing physical phenomena in ultra-thin oxide superlattices where the confinement effects are prominent. For instance, the high-temperature superconductivity was predicted in confined nickelate layers,\cite{Chaloupka:Prl,Hansmann:Prl} and the spin and orbital configurations of LaTiO$_3$ (LTO) can be effectively manipulated when confined within the LaAlO$_3$ barriers.\cite{Seo:Prl,Lee:Prb14} In these ultra-thin oxide heterostructures, reconstructions always play critical roles, which include both the electronic and lattice reconstructions induced by various physical reasons like charge leakage,\cite{Dong:Prb08.3} epitaxial strain,\cite{Dong:Prb12,Zhang:Prb12} or exchange interaction. Just due to these reconstructions, oxide heterostructures often exhibit interesting phenomena and fascinating properties which are absent in individual material.\cite{Zubko:Arcmp}

Besides aforementioned several mechanisms, the tailoring of structural orientation is also an effective route to tune the physical properties in oxide heterostructures. Recent experimental and theoretical studies both suggest that the magnetism and electronic structure of transition metal oxides may change dramatically when their structural orientation is rotated. For example, LaFeO$_3$-LaCrO$_3$ SLs can show different magnetism when growing along different crystal axes,\cite{Ueda:Sci,Ueda:Jap,Zhu:Jap} and exchange bias was observed in the ($111$)-oriented LaNiO$_3$/LaMnO$_3$ SLs but not found in the ($001$)-oriented one.\cite{Dong:Prb13,Gibert:Nm} Furthermore, topological phases were predicted to emerge in the ($111$)-oriented LaNiO$_3$ or other oxide bilayers.\cite{Ruegg:Prb12,Yang:Prb,Xiao:Nc}

\begin{figure}
\centering
\includegraphics[width=0.48\textwidth]{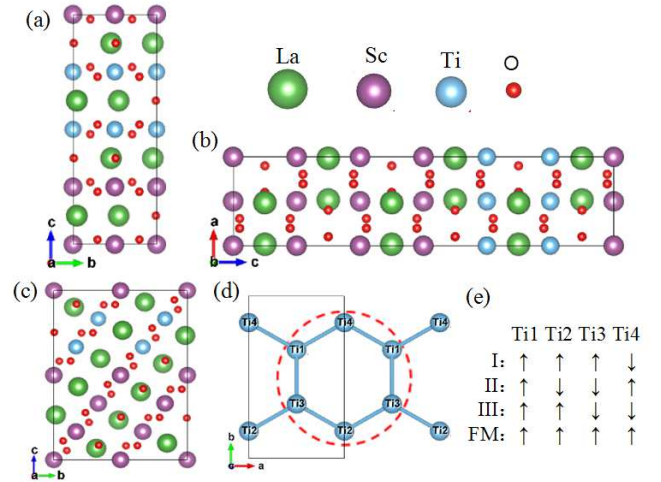}
\caption{(Color online) Structures of the LSO-LTO SLs. (a) The ($001$)-oriented (LSO)$_2$/(LTO)$_2$. (b) The ($001$)-oriented (LSO)$_4$/(LTO)$_2$. (c-d) The ($111$)-oriented (LSO)$_4$/(LTO)$_2$: (c) side view; (d) top view along the ($111$) direction. In (d), each unit cell includes four Ti ions (marked as Ti$1$, Ti$2$, Ti$3$, and Ti$4$). (e) Sketch of four magnetic orders in the ($111$)-bilayer. Here the type-I is ferrimagnetic and both the type-II and type-III are antiferromagnetic.}
\label{diagram}
\end{figure}

In this work, the magnetism and electronic structure of ($001$)- and ($111$)-oriented LTO bilayers have been studied. LaScO$_3$ (LSO) is adopted as the barrier layer and substrate. The crystal structures of such (LSO)$_n$/(LTO)$_2$ SLs are shown in Fig.~\ref{diagram}(a-c), where the subscripts $n$ and $2$ denote the unit layers within one period.

LTO bulk is a G-type antiferromagnetic (G-AFM) Mott insulator with weak GdFeO$_3$-type distortion.\cite{Mochizuki:Njp} It has the orthorhombic perovskite structure (space group \textit{Pbnm})\cite{Cwik:Prb} and the experimental lattice constants are $a$=$5.636$ \AA{}, $b$=$5.618$ \AA{}, $c$=$7.916$ \AA{}.\cite{Komarek:Prb} One crystal unit cell contains four chemical formula units. There is one electron occupying Ti's $3d$ $t_{\rm 2g}$ orbitals. The LSO is a nonmagnetic band insulator with the $d^0$ (Sc$^{3+}$) electronic configuration. Its optical band gap, opened between the occupied O's $2p$ bands and unoccupied Sc's $3d$ bands, is about $6.0$ eV.\cite{He:Prb12} LSO also forms the orthorhombic perovskite structure (space group \textit{Pbnm}) with the lattice constants as $a$=$5.678$ \AA{}, $b$=$5.787$ \AA{}, $c$=$8.098$ \AA{}.\cite{Geller:Ac}

In the present work, we study the (LSO)$_n$/(LTO)$_2$ SLs for the following physical considerations. First, their lattice constants are proximate, which ensure the possibility for epitaxial growth in experiments and thus make our theoretical prediction realistically meaningful. Second, the A-site cations are both La for these two materials, which make a unique termination and exclude the complex polar catastrophe due to the polar discontinuity in some SLs.\cite{Nakagawa:Nm} Third, since LSO is highly insulating with a very large band gap, it may paly as a good barrier to confine the $3d$ electrons within the quantum well of Ti layers, which reshapes the structural dimension of LTO. Last but not the least, the tensile strain from LSO will tune the crystal structure of LTO.\cite{Weng:Jap14}

Our density-functional theory (DFT) calculations were performed using the projector-augmented wave (PAW) potentials, as implemented in the Vienna \textit{ab} initio Simulation Package (VASP).\cite{Kresse:Prb,Kresse:Prb96} The electronic correlation is treated using the local density approximation (LDA) with Hubbard $U$. The cutoff energy of plane-wave is $500$ eV and the on-site Hubbard interaction is set as $U-J$=$2.3$ eV for Ti's $3d$ electrons and $8$ eV for La's $4f$ electrons using the Dudarev implementation.\cite{Dudarev:Prb}

First, the magnetic ground state and electronic structure of LTO bulk are calculated. Both the lattice parameters and internal positions of ions are fully relaxed from the experimental structure and with the experimental magnetic order (G-AFM). The optimized structure gives $a$=$5.634$ \AA{}, $b$=$5.551$ \AA{}, $c$=$7.868$ \AA{}, which are close to the experimental values mentioned before. To check the magnetic ground state, the total energies of ferromagnetic (FM), A-type antiferromagnetic (A-AFM), C-type antiferromagnetic (C-AFM), and G-AFM states are calculated. In our LDA+$U$ calculation, it is demonstrated that the G-AFM has the lowest energy and the local magnetic moment is $0.683$  $\mu_{\rm B}$/per Ti, as shown in Table~\ref{table1}, which are also in agreement with the experimental results.\cite{Cwik:Prb} In addition, the insulating behavior with an energy gap of $0.26$ eV is very close to experimental value $0.2$ eV.\cite{Okimoto:Prb} In short, the results of our DFT calculation agree well with the experimental data, guaranteeing the validity of theoretical method and parameters.

Subsequently, the magnetism and electronic structure of the (LSO)$_n$/(LTO)$_2$ ($n$=$2$, $4$ for the ($001$)-oriented and $4$ for the ($111$)-oriented) SLs are studied. Our previous DFT calculation reported that the LTO film has a tendency to approach the C-AFM under tensile strain from substrates with relative large lattice constants, such as LSO used in the present work.\cite{Weng:Jap14} For these SLs, both the lattice constants along the $c$-axis and inner atomic positions are fully optimized until the Hellman-Feynman forces are less than $10$ meV/\AA{}.

For the ($001$) cases, four magnetic orders (FM, A-AFM , C-AFM, and G-AFM) are adopted for comparison. The optimized equilibrium values for lattice constant along the $c$-axis are $15.67$ \AA{} and $23.68$ \AA{} for the (LSO)$_2$/(LTO)$_2$ and (LSO)$_4$/(LTO)$_2$, respectively. According to the DFT results summarized in Table~\ref{table1}, the G-AFM have the lowest energy and the C-AFM is the first excited state with a proximate energy due to the strain effect, similar to the pure LTO film grown on the LSO substrate.\cite{Weng:Jap14} The energy difference $\Delta E$ and the local magnetic moment $M_{\rm Ti}$ do not change too much when $n$ changes from $2$ to $4$ for these ($001$)-oriented SLs. Such a behavior is due to the highly insulating LSO barrier, namely these ultra-thin LSO layers are enough to isolate LTO bilayers.

\begin{table}
\footnotesize
\caption{Summary of DFT results. Various magnetic orders are calculated. The G-AFM or type-III AFM are taken as the reference state for energy comparison. The energy difference $\Delta E$ is in unit of meV/Ti. The magnetic moment $M_{\rm Ti}$ is in unit of $\mu_B$/Ti, obtained within the Wigner-Seitz sphere as specified by VASP. The band gap is in unit of eV.}
\centering
\begin{tabular*}{0.48\textwidth}{@{\extracolsep{\fill}}lllllr}
\hline \hline
\ & FM & A-AFM (I) & C-AFM (II) & G-AFM (III)\\
\ &$\Delta E|M_{\rm Ti}$ & $\Delta E|M_{\rm Ti}$  & $\Delta E|M_{\rm Ti}$& $M_{\rm Ti}|$gap\\
\hline
LTO bulk & $17.2|0.825$  & $1.5|0.747$ & $17.3|0.726$ & $0.683|0.26$\\
$n=2$ ($001$) & $6.5|0.821$ & $7.5|0.796$ & $0.5|0.752$ & $0.739|0.62$\\
$n=4$ ($001$) & $6.4|0.822$ & $7.4|0.797$ & $0.2|0.752$ & $0.740|0.62$\\
$n=4$ ($111$) & $11.1|0.804$ & $4.9|0.776$ & $2.2|0.781$ & $0.752|0.60$\\
\hline \hline
\end{tabular*}
\label{table1}
\end{table}

Different from the ($001$) cases, the ($111$) bilayer forms the buckled honeycomb lattice, as shown in Fig.~\ref{diagram}(d). In such a SL, each unit cell includes four Ti ions, marked as Ti$1$, Ti$2$, Ti$3$, and Ti$4$ in Fig.~\ref{diagram}(d). Along the ($111$) axis, Ti$1$ and Ti$2$ belong to the first layer, while Ti$3$ and Ti$4$ are located in the second layer. The most nontrivial physics is that the coordination numbers of Ti-O-Ti bonds are reduced to three, while the numbers of ($001$) bilayer are five.\cite{Dong:Prb13} To determine the ground state, several possible magnetic orders have been calculated, as sketched in Fig.~\ref{diagram}(e). By comparing the total energies of these magnetic orders, it is found that the type-III AFM has the lowest energy, in which all nearest-neighbor spin pairs are antiparallel, similar to the G-AFM in ($001$)-oriented SLs. In this sense, the spin correlation is not altered when the orientation is rotated from the ($001$) direction to the ($111$) one.

\begin{figure}
\centering
\includegraphics[width=0.5\textwidth]{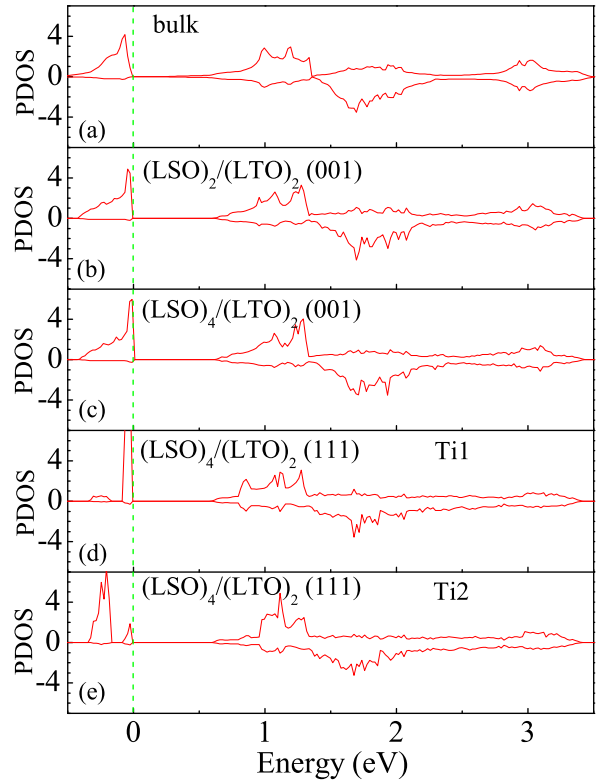}
\caption{(Color online) The PDOS of majority-spin Ti ions. (a) In bulk; (b) In the ($001$) $n=2$ SL; (c) In the ($001$) $n=4$ SL; (d-e) In the ($111$) $n=4$ SL.
}
\label{dos}
\end{figure}

Although the ground state magnetism has no apparent change in both SLs along the ($001$) and ($111$) directions compared with the LTO bulk and pure film, the electronic structure shows a non-trivial difference near the Fermi level. As shown in Fig.~\ref{dos}, the projected density of states (PDOS) of Ti ions near the Fermi level are significantly narrowed and split into two bands in the ($111$) SL, while it is almost identical among the bulk and the ($001$) SLs. In fact, for ($111$) SL, the oxygen octahedral distortions exhibit non-negligible difference between Ti$1$ and Ti$2$ by considering the Ti-O bond lengths and Ti-O-Ti bond angles. And this symmetry breaking originates from the incompatibility between the threefold rotation symmetry of the cubic [$111$] axis and the twofold rotation symmetry of orthorhombic structure. The band gaps of bulk and SLs are also summarized in Table.~\ref{table1}. Both the ($001$) and ($111$) SLs show much larger gaps comparing with the bulk.

\begin{figure}
\centering
\includegraphics[width=0.5\textwidth]{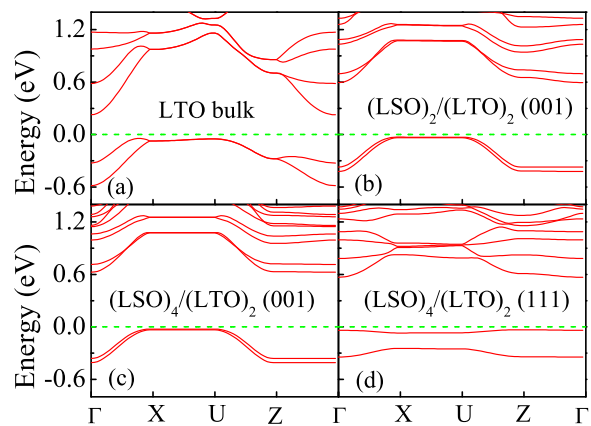}
\caption{(Color online) The majority-spin band structure. (a) The LTO bulk; (b) The $n=2$ ($001$) SL; (c) The $n=4$ ($001$) SL; (d) The $n=4$ ($111$) SL.}
\label{band}
\end{figure}

The narrowing of occupied bands in the ($111$) SL can be also revealed in the band structures. As shown in Fig.~\ref{band}, the ($001$) cases show broad occupied $t_{\rm 2g}$ bands: $0.39$ eV and $0.38$ eV in the $n=2$ and $n=4$ SLs, respectively. These values are lower than the bulk value ($0.54$ eV), implying a confinement effect. However, the occupied states are split into two nearly flat bands in the (111) case, which's widths are only $0.04$ eV and $0.1$ eV, significantly lower than the ($001$) values. It is well known that the width of a band is in proportional to the kinetic energy of electron. Then, the $t_{\rm 2g}$ electrons in the ($001$) SLs are more itinerant, while they are seriously confined in the ($111$) one. Similar confinement effects have also been revealed in other ultra-thin oxide superlattices.\cite{Chaloupka:Prl,Hansmann:Prl,Seo:Prl,Lee:Prb14} Thus, the Mott-insulator transition, determined by the ratio between Hubbard $U$ and bandwidth, will become easier in the confined electron systems. However, the band gaps, shown in Fig.~\ref{dos}, Fig.~\ref{band}, and summarized in Table~\ref{table1}, show an even slightly larger value in the ($001$) SLs than the ($111$) one. The physical reason is that the band gap here is between the occupied $t_{\rm 2g}$ singlet and the empty $t_{\rm 2g}$ doublets due to the structural distortions (e.g. the Jahn-Teller distortions) while the upper Hubbard bands are even much higher in energy.

According to the energy differences among various magnetic orders, the superexchange coefficients can be obtained by mapping the system to a classical Heisenberg model with normalized spins ($|S|=1$). Here, the in-plane exchange $J_{ab}$ and out-of-plane exchange $J_c$ of LTO are calculated in the ($001$) cases, while the nearest-neighbor exchange $J_1$ and next-nearest-neighbor exchange $J_2$ are accounted in the ($111$) case. Our calculations find that $J_{ab}$=$1.5$ meV and $J_c$=$0.5$ meV in the ($001$) (LSO)$_2$/(LTO)$_2$, and $J_{ab}$=$1.6$ meV and $J_c$=$0.2$ meV in the ($001$) (LSO)$_4$/(LTO)$_2$, as summarized in Table~\ref{table2}. It is clearly that the in-plane exchange does not change too much with increase $n$, as expected. For the ($111$) SL, $J_1=4.9$ meV and $J_2=1.9$ meV. First, the nearest-neighbor exchange $J_1$ in the ($111$) case is much stronger than both $J_{ab}$ and $J_c$ in the ($001$) cases, which will certainly enhance the N\'{e}el temperature significantly. In addition, the AFM $J_2$ will bring weak exchange frustration into this honeycomb lattice.

\begin{table}
\caption{Exchange coefficients calculated by mapping the DFT energies to a classical spin model. For the ($001$)-oriented SLs, the in-plane exchange $J_{ab}$ and out-of-plane exchange $J_c$ are calculated, while the nearest-neighbor exchange $J_1$ and next-nearest-neighbor exchange $J_2$ are calculated for the ($111$) case.}
\centering
\begin{tabular*}{0.48\textwidth}{@{\extracolsep{\fill}}llllr}
\hline \hline
\centering
Exchange & ($001$) & ($001$) & ($111$)\\
coupling & $n$=$2$ & $n$=$4$ & $n$=$4$\\
\hline
$J_{ab}$ ($J_1$) & $1.5$ & $1.6$ & $4.9$\\
$J_c$ ($J_2$)	& $0.5$	& $0.2$	& $1.9$\\
\hline \hline
\end{tabular*}
\label{table2}
\end{table}

\begin{figure}
\centering
\includegraphics[width=0.5\textwidth]{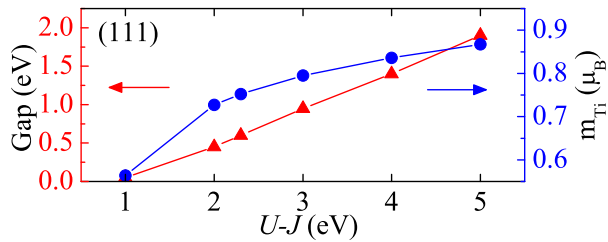}
\caption{(Color online) The band gap (left axis) and local moment of each Ti (right axis) as a function of the Hubbard coupling strength applied on Ti's $3d$ electrons in the $n$=$4$ ($111$) SL.}
\label{Gap}
\end{figure}

Moreover, according to Table~\ref{table1} or band structures (Fig.~\ref{dos} and Fig.~\ref{band}), it is found that the insulating behavior is still preserved in both the ($001$)- and ($111$)-oriented SLs, and the band gaps become even larger compared with the LTO bulk. The reason may be the enhanced lattice distortions originated from the lattice mismatch between LTO and LSO and reduced dimension. Of course, the Hubbard replusion is always a crucial driven force to open band gaps in Mott insulators. Here, the LDA+$U$ method is tested for the ($111$)-oriented SL by tuning the value of $U-J$ from $1$ eV to $5$ eV. As shown in Fig.~\ref{Gap}, both the band gap and local moment increase with the effective Hubbard parameter $U-J$. And the band gap tends to close when $U-J$ is equal to or below $1$ eV.

In summary, magnetism and electronic structure of the ($001$)- and ($111$)-oriented LSO-LTO superlattices have been studied using the LDA+$U$ method. Although the ground state of LTO bilayer remains the G-type antiferromagnetic insulator in these heterostructures, electronic structures show a non-trivial difference near the Fermi level due to the different dimensional confinement. Furthermore, the exchange couplings are also calculated, which suggests a possible great enhancement of N\'{e}el temperature in the ($111$)-oriented superlattice.

Work was supported by the $973$ Projects of China (Grant No. 2011CB922101), NSFC (Grant Nos. 11274060 and 51322206).

\bibliographystyle{apsrev4-1}
\bibliography{ref}
\end{document}